\documentclass[a4paper]{jpconf}
\usepackage{graphicx}
\begin{document}
\title{Semiclassical and quantum Liouville theory}

\author{Pietro Menotti}

\address{Department of Physics, University of Pisa and INFN Sezione di
Pisa, Italy} 

\ead{menotti@df.unipi.it}

\begin{abstract}
We develop a functional integral approach to quantum Liouville field
theory completely independent of the hamiltonian approach. To this end
on the sphere topology we solve the Riemann-Hilbert problem for three
singularities of finite strength and a fourth one infinitesimal, by
determining perturbatively the Poincar\'e accessory parameters. This
provides the semiclassical four point vertex function with three
finite charges and a fourth infinitesimal. Some of the results are
extended to the case of $n$ finite charges and $m$ infinitesimal.
With the same technique we compute the exact Green function on the
sphere on the background of three finite singularities. Turning to the
full quantum problem we address the calculation of the quantum
determinant on the background of three finite charges and of the
further perturbative corrections. The zeta function regularization
provides
a theory which is not invariant under local conformal
transformations. Instead by employing a regularization suggested in
the case of the pseudosphere by Zamolodchikov and Zamolodchikov we
obtain the correct quantum conformal dimensions from the one loop
calculation and we show explicitly that the two loop corrections do
not change such dimensions. We then apply the method to the case of
the pseudosphere with one finite singularity and compute the exact
value for the quantum determinant. Such results are compared to those
of the conformal bootstrap approach finding complete agreement.
\end{abstract} 

\section{Introduction}
Liouville theory \cite{CT,DJ,DFJ,GN} plays an important role in several
branches of 
physics and mathematics. It is deeply related to the problem of
uniformization of Riemann surfaces; it plays a major role in 2+1
dimensional gravity also in presence of matter; it appears in two
dimensional gravity as an outcome of the conformal anomaly, in non
critical string theory, in special models of 2D critical string
theory and in the AdS-CFT correspondence \cite{GKP} and 
recently in some D-brane models \cite{maldacena}.

There are practically three approaches to the quantum problem: the
hamiltonian, the functional and the conformal bootstrap approach. 

The hamiltonian approach (Minkowski space) was first pursued by
Curtright and Thorn \cite{CT} and by D'Hoker, Friedman and Jackiw
\cite{DJ,DFJ}. In papers \cite{DJ,DFJ} it was shown that in presence
of a ground state the theory develops a spontaneous symmetry breaking
of Poincar\'e invariance and in particular of translational
invariance. As a result the theory is defined on the half-line and of
the whole conformal group, present in the classical action, only the
subgroup $SO(2,1)$ survives. In \cite{CT} (see also \cite{JW,KN}) the
theory is compactified on a circle, there is no ground state, but by
properly modifying the energy momentum tensor it is possible to show
that the emerging quantum theory is invariant under the whole
conformal group. The central charge of the theory turns out to be
$c=1+6 Q^2$, where $Q=1/b+b$ being $b$ the coupling constant; the
dimension of the vertex function $V_\alpha = \exp(2\alpha \phi)$ is
$\Delta_\alpha=\alpha(Q-\alpha)$.

The functional approach \cite{GL,teschnerrev,ZZsphere,ZZpseudosphere}
has been used mostly at the formal level in combination with the
conformal bootstrap approach. One accepts a priori the invariance of
the theory under the full conformal group. Some results are borrowed
from the hamiltonian approach; the integral computed at some special
point in the space of the charges and interpolation formulae devised
to extend the treatment to the general case. Four point functions on
the sphere (or two point function on the pseudosphere) are computed
when one of the vertex function is a degenerate field (Teschner
trick). Consistency with the formulation in the crossed channel
originates some difference equation. After imposing a further symmetry
on the result one gets the final answer.

The most important results are the exact expressions for the three
point function on the sphere \cite{dornotto,ZZsphere,teschnerrev}, the
one point function on the pseudosphere (ZZ-brane)
\cite{ZZpseudosphere}, and several similar results in the case of
boundary conformal Liouville theory (FZZT-brane) \cite{FZZ,teschnerbound}.

The subject of this talk is to widen the range of applications of the
standard functional approach. By this we mean the formulation in which
one first computes a stable background and then integrates over the
fluctuations around it. We shall see that it is possible to develop
techniques which allow the resummation of infinite classes of Feynman
graphs and compare such results with formulas derived in the conformal
bootstrap approach. There is a more general aspect in this kind of
research. It is well know that a quantum field theory is characterized
non only by an action but also by a regularization and
renormalization procedure. Thus there is a non trivial question to
answer i.e.: Which is the correct action to start with and which
regularization procedure has one to adopt in order the produce
perturbatively result which are consistent with invariance under the
full conformal group? As we shall  see not all regularization procedures
give rise to a field theory satisfying such requirements.

\section{Classical Liouville theory}\label{clt} 
We start with the sphere topology. The regularized classical action
for the Liouville theory on the sphere in presence of $N$ sources is given by
\cite{ZZsphere} 
\begin{equation}\label{regularizedaction}
S_L[\phi] = \lim_{\stackrel{\varepsilon_n \rightarrow 0}{R \rightarrow
\infty}} \Bigg\{ \int_{\Gamma_{\varepsilon ,R}} \left[ \frac{1}{\pi}
\partial_{z} \phi \partial_{\bar{z}} \phi + \mu e^{2b\phi} \right] \;
i \frac{dz \wedge d\bar{z}}{2} 
\end{equation}
$$
+\frac{Q}{2\pi i}
\oint_{\partial\Gamma_{R}} \phi \, \left( \frac{dz}{z} -
\frac{d\bar{z}}{\bar{z}} \right) + Q^2 \, \log R^2 
-\frac{1}{2\pi i} \sum_{n=1}^{N} \alpha_n \,
\oint_{\partial\Gamma_{n}} \phi \, \left( \frac{dz}{z-z_n} -
\frac{d\bar{z}}{\bar{z}-\bar{z}_n} \right) - \sum_{n=1}^{N} \alpha_n^2
\, \log \varepsilon_n^2 \Bigg\}
$$
where $z_n$ and $\alpha_n$ are position and charge of the $n$-th
source. The domain of integration is the region $\Gamma_{\varepsilon,R} =
\left\{ |z| < R\right\} \setminus \bigcup_{n} \left\{ |z-z_n|< \varepsilon_n
\right\}$, $\partial\Gamma_R$ is the border around infinity while
$\partial \Gamma_{n}$ is the border around the $n$-th source. Here $Q$
is a parameter linked to the transformation law of the Liouville
field. Classically its value is $Q=1/b$.

In order to examine the semiclassical limit of the $N$-point function
it is useful \cite{ZZsphere} to go over to the field $\varphi = 2b
\phi$. The corresponding charges are $\eta_n = \alpha_n b$ and 
the action takes the form
\begin{equation}
S[\varphi] = b^2 S_L[\phi] = \lim_{\stackrel{\varepsilon_n
\rightarrow 0}{R \rightarrow 
\infty}}
\int_{\Gamma_{\varepsilon,R}} \left[
			\frac{1}{4 \pi} \partial_{z} \varphi
			\partial_{\bar{z}} \varphi + b^2 \mu e^{\varphi}
			\right] \; i \frac{dz \wedge d\bar{z}}{2} 
\end{equation}
$$
+ \frac{bQ}{4\pi i}
			\oint_{\partial\Gamma_{R}} \varphi \left(
			\frac{dz}{z} - \frac{d\bar{z}}{\bar{z}}
			\right) + (bQ)^2 \, \log R^2 
			  -\frac{1}{4\pi i} \sum_{n=1}^{N} \eta_n \,
			\oint_{\partial\Gamma_{n}} \varphi \, \left(
			\frac{dz}{z-z_n} -
			\frac{d\bar{z}}{\bar{z}-\bar{z}_n} \right) -
			\sum_{n=1}^{N} \eta_n^2 \, \log
			\varepsilon_n^2.
$$
The field $\varphi$ behaves like
\begin{equation}\label{eq:asymptotics}
\left\{
\begin{array}{l}
\varphi(z) = - 2\eta_n \, \log |z-z_n|^2 + O(1) \qquad {\rm for }~
z \rightarrow z_n \\ 
\varphi(z) = - 2b Q \, \log |z|^2 + O(1)
\qquad \qquad {\rm for }~ z \rightarrow \infty.
\end{array}
\right.
\end{equation}
We decompose the field $\varphi$ into the sum of a classical background
$\varphi_B$ and a quantum fluctuation $\varphi =
\varphi_B + 2b\chi$. The action becomes
$
S_L[\varphi_B,\chi] = S_{cl}[\varphi_B]+ S_q[\varphi_B,\chi]
$
where
\begin{equation}\label{classicalaction}
S_{cl}[\varphi_B] = \lim_{\stackrel{\varepsilon_n \rightarrow 0}{R
\rightarrow 
\infty}}\frac{1}{b^2}\left[
\frac{1}{8\pi}\int_\Gamma \left(\frac{1}{2}(\partial_a\varphi_B)^2 +8\pi\mu
b^2e^{\varphi_B}\right)d^2 z \right.
\end{equation}
$$
-\sum_{n=1}^{N} \left(\eta_n \,
			\frac{1}{4\pi i}\oint_{\partial\Gamma_{n}}
			\varphi_B \,  
			(\frac{dz}{z-z_n}-\frac{d\bar z}{\bar z-\bar z_n})
+\eta_n^2 \log\varepsilon_n^2\right) \nonumber\\
\left.+\frac{1}{4\pi i}\oint_{\partial\Gamma_{R}} \varphi_B \, \left(
			\frac{dz}{z} - \frac{d\bar{z}}{\bar{z}}
			\right) + \, \log R^2\right] 
$$
and
\begin{eqnarray}\label{quantumaction}
S_q[\varphi_B,\chi] &=& \lim_{\stackrel{\varepsilon_n \rightarrow 0}{R
\rightarrow 
\infty}} \frac{1}{4\pi}\int_\Gamma\left((\partial_a
\chi)^2 + 4\pi \mu e^{\varphi_B}(e^{2b\chi}-1-2b\chi)\right)d^2z
\nonumber\\
&+& (2+b^2)\ln R^2 +
\frac{1}{4\pi
  i}\oint_{\partial\Gamma_R}\varphi_B\left(\frac{dz}{z}-\frac{d\bar z}{\bar
  z}\right)+ 
\frac{b}{2\pi i}\oint_{\partial\Gamma_R}\chi \left(\frac{dz}{z}-\frac{d\bar
  z}{\bar z}\right). 
\end{eqnarray}
The terms in the second row arise from having chosen $Q=1/b+b$.  The
main difficulty with the sphere topology is that there is no stable
solution to the Liouville equation derived from the classical action
in absence of sources. This is due to some inequalities which go back to
Picard \cite{picard}, i.e. we must have $\eta_n\leq 1/2$ and $\sum
\eta_n > 1$ which implies that at least three singularities have to be
present and 
sufficiently strong. Thus there cannot be a usual perturbative
expansion on the sphere with weak sources, unless one uses the fixed
area approach; but functionally integrating with constraints is more
difficult. 

From the classical action (\ref{classicalaction}) one derives
the Liouville equation
\begin{equation}\label{classicalequation}
-\Delta \varphi + 8\pi\mu b^2 \, e^{\varphi} = 8\pi \sum_{n=1}^{N}
 \eta_n \delta^2(z-z_n)
\end{equation}
whose solutions can be reduced to the solution
of the fuchsian equation 
\begin{equation}\label{fuchsian}
y''(z) + Q(z)y(z) = 0
\end{equation}
where
\begin{equation}\label{eqf}
Q(z) = \sum_{n=1}^N \left(\frac{1-\lambda_n^2}{4(z-z_n)^2} +
\frac{\beta_n}{2(z-z_n)} \right).
\end{equation}
Here in addition to the parameters $\lambda_n = 1-2\eta_n$ related to
the charges, also the Poincar\'e accessory parameters $\beta_n$
appear. These accessory parameters must satisfy three constraints known as
Fuchs relations
which in the case of only three singularities are sufficient to
determine the three accessory parameters. In the case of more that
three singularities the accessory parameters have to be fixed by
imposing the monodromy of the conformal factor $\varphi$.

The solution of equation (\ref{classicalequation}) is given by
\begin{equation}\label{eq:conformal-factor}
e^{\varphi_c} = \frac{1}{\pi\mu b^2}\,\frac{|w_{12}|^2}{{\left(y_2
\bar{y_2} - y_1 \bar{y_1} \right)}^2}
\end{equation}
where $w_{12}=y_1 y'_2 - y_1'y_2$ is the constant wronskian and the
two solutions  $y_1$ and $y_2$ of (\ref{fuchsian}) must be
chosen in such a way that their monodromy group is $SU(1,1)$ in order
to ensure that the Liouville field $\varphi(z)$ is one-valued on the
whole complex plane.  In the case of only three singularities the
conformal factor is given in terms of hypergeometric functions.

The classical action (\ref{classicalaction}) has very simple
transformation properties under 
$SL(2,C)$ \cite{takhtajan,MT2}. Moreover on the solution of Liouville
equation it satisfies two 
important relations. The first is easily derived from the form of the
action and reads
\begin{equation}\label{Xrelation}
\frac{\partial S_{cl}}{\partial \eta_i} = - X_i
\end{equation}
where $X_i$ is the finite part of the field $\varphi_c$ at $z_i$
\begin{equation} 
\varphi_c(z) = -2\eta_i\log|z-z_i|^2 +X_i+ o(|z-z_i|).
\end{equation} 
The second relation is the so called Polyakov relation
\cite{ZT1,CMS1,ZT2}
\begin{equation}\label{polyakovrelation}
\frac{\partial S_{cl}}{\partial z_i} = -\frac{\beta_i}{2}
\end{equation}
which directly relates the accessory parameters to the classical
Liouville action. These two relations properly rewritten and
interpreted, contain all the hamiltonian structure of 2+1 dimensional
gravity \cite{CMS2,CM}. 
Using relations (\ref{Xrelation},\ref{polyakovrelation}) it is
possible to compute the semiclassical 
limit of the three-point function, which is related
to the value of the classical action. 
Integrating the differential system
(\ref{Xrelation},\ref{polyakovrelation}) one obtains \cite{ZZsphere}
\begin{eqnarray}\label{classicalthreepoint}
S_{cl}[z_1,z_2,z_3;\eta_1,\eta_2,\eta_3] & = &
			(\delta_1+\delta_2-\delta_3)\log|z_1-z_2|^2  
			+ (\delta_2+\delta_3-\delta_1)\log|z_2-z_3|^2
			\nonumber\\ & &
			(\delta_3+\delta_1-\delta_2)\log|z_3-z_1|^2 +
			S_{cl}[0,1,\infty;\eta_1,\eta_2,\eta_3]
\end{eqnarray}
where $\delta_i = \eta_i ( 1 - \eta_i)$ are the semiclassical
dimensions and
\begin{eqnarray}
S_{cl}[0,1,\infty;\eta_1,\eta_2,\eta_3] &=& S_0 +
		\left(\eta_1+\eta_2+\eta_3 - \frac{3}{2}\right)\log
		(\pi\mu b^2) + 3F(1) \nonumber\\ & &
		-F(2\eta_1)-F(2\eta_2)-F(2\eta_3) +
		F(\eta_1+\eta_2+\eta_3-1) \nonumber\\ & &
		+F(\eta_3+\eta_2-\eta_1) + F(\eta_2+\eta_1-\eta_3) +
		F(\eta_3+\eta_1-\eta_2).
\end{eqnarray}
The function $F$ is given by
\begin{equation}
F(x) = \int_{1/2}^{x} \, \log \gamma(s) \, ds.
\end{equation}
where as usual $\gamma(x) = \Gamma(x)/\Gamma(1-x)$.
Notice that a regulation procedure of the action is necessary also
at the classical level and it gives rise to the semiclassical
dimension $\delta_i = \eta_i ( 1 - \eta_i)$. These are not yet the
quantum dimensions obtained in the hamiltonian approach.
To proceed one needs the Green function on the background of three
(not small) sources. 

\section{The semiclassical four point function}

In this section we shall determine the classical action in presence of
three finite singularities and a fourth infinitesimal; such a calculation
gives the semiclassical four point function for vertices with
three finite charges and the fourth small \cite{MV}.
The procedure we shall use in presence of a fourth weak singularity is to
solve perturbatively the fuchsian equation associated to the Liouville
equation leaving the fourth small accessory parameter $\beta_4$ free,
and then determine it by imposing the monodromy condition on the
conformal factor.
Given four singularities, by means of an $SL(2,C)$ transformation we
can take  three of them in $0,1,\infty$. The position of the fourth will
be called $t$ and the coefficient $Q$ in the fuchsian equation
becomes
\begin{equation}
Q(z) = \frac{1-\lambda_1^2}{4z^2} + \frac{1-\lambda_2^2}{4(z-1)^2} +
	\frac{1-\lambda_4^2}{4(z-t)^2} + \frac{\beta_1}{2z} +
	\frac{\beta_2}{2(z-1)} + \frac{\beta_4}{2(z-t)}~.
\end{equation}
For the source in $t$ of infinitesimal strength we shall write
$\lambda_4 = 1-2\varepsilon$ and $\beta_4 = \varepsilon \beta$ and our
aim will be to determine $\beta$. Using Fuchs relations we have
\begin{eqnarray}
\beta_1 & = & \frac{1-\lambda_1^2-\lambda_2^2+\lambda_3^2}{2} +
\varepsilon \, \left[ (t-1)\beta +2\right] + O(\varepsilon^2) \nonumber\\
\beta_2 & = & -\frac{1-\lambda_1^2-\lambda_2^2+\lambda_3^2}{2} -
\varepsilon \, \left[ 2+t\beta \right] + O(\varepsilon^2)
\end{eqnarray}
and we write
\begin{equation}
Q(z) = Q_0(z) + \varepsilon \, q(z)
\end{equation}
where $Q_0(z)$ stays for the coefficient of the three singularity
problem, while $q(z)$ is the perturbation
\begin{equation}\label{eq:q}
q(z) = \frac{1}{2}\left[ \frac{(t-1)\beta+2}{z} - \frac{2+t\beta}{z-1}
+ \frac{\beta}{z-t} + \frac{2}{(z-t)^2} \right].
\end{equation}
After writing $y = y_0 +\varepsilon\delta y$, being $y_0$ a solution of the
unperturbed equation, we have to first order in $\varepsilon$ the
inhomogeneous equation
\begin{equation}\label{inhomogeneous}
(\delta y)'' + Q\, \delta y = -q \, y_0~.
\end{equation}
Such an equation can be solved by a well known method and
in our case we have
\begin{equation}\label{eq:soluzione-pertubativa}
\delta y_i = -\frac{1}{w_{12}} \int_{z_0}^{z} dx \; \left[ y_1(x) y_2(z) -
y_1(z) y_2(x) \right] \, q(x) y_i(x)
\end{equation}
being $w_{12}= y_1 \,y_2'- y_1'\,y_2$ the constant
wronskian and $z_0$ an
arbitrary base point in the complex plane.  
It will be useful to define the following integrals
\begin{equation}\label{eq:integrali}
I_{ij}(z) \equiv \int_{z_0}^z dx \; y_i(x) y_j(x) \, q(x).
\end{equation}
We must now compute the monodromy matrices around $0,1,t$ and impose
on them the $SU(1,1)$ nature. This will determine uniquely the
parameter $\beta$.
The calculation gives \cite{MV}
\begin{equation}\label{beta}
\beta = -4\frac{\kappa \, \bar{y}_1 y_1' - \bar{y}_2 y_2'}{\kappa \,
\bar{y}_1 y_1 - \bar{y}_2 y_2}(t)=2 \partial_z\varphi^0_c(z)|_{z=t}
\end{equation}
being $\kappa= |k_0|^4$ with $k_0$ the parameter which enters the
three-singularity conformal factor
\begin{equation}
e^{2b\phi_c^0}= \frac{1}{\pi\mu b^2} \,
\frac{{w_{12}}^2}{({|k_0|^2y_1 \bar{y}_1 - |k_0|^{-2} y_2 \bar{y}_2
)}^2} 
\end{equation}
and
\begin{equation}\label{eq:soluzione-fattore-conforme}
e^{2b\phi_c}=e^{\varphi_c} = \frac{1}{\pi\mu b^2} \,
\frac{{w_{12}}^2}{{\left(Z_1 \bar{Z}_1 - Z_2
\bar{Z}_2\right)}^2}
\end{equation}
with 
\begin{eqnarray}
Z_1(z) & = & k_0 \, \left[ \left( 1 + \varepsilon \frac{I_{12}(z) +
    h}{w_{12}}\right)\, y_1(z) - \varepsilon \frac{I_{11}(z)}{w_{12}}\, y_2(z)
  \right] \label{eq:soluzione-monodroma-1} \nonumber\\ Z_2(z) & = &
\frac{1}{k_0} \, \left[ \varepsilon \frac{I_{22}(z)}{w_{12}}\, y_1(z) +
  \left( 1 - \varepsilon \frac{I_{12}(z) + h}{w_{12}}\right)\, y_2(z)
  \right] \label{eq:soluzione-monodroma-2}
\end{eqnarray}
where the $h$ can also be computed \cite{MV}.
The functions $Z_1,Z_2$ have $SU(1,1)$ monodromies around all
singularities and as such determine a globally monodromic conformal
factor satisfying the Liouville equation.
We can now compute the conformal factor in presence of our four
sources to first order in $\varepsilon$ 
\begin{eqnarray}\label{expphi}
e^{\varphi_c} & = & e^{\varphi_c^0}\left\{ 1 -\varepsilon\, \frac{2}{w_{12}
					\left(\kappa y_1 \bar{y}_1 -
					y_2 \bar{y}_2\right)}
					\right. \\ & &
					\left. \qquad\qquad \left[
					\left(\kappa y_1 \bar{y}_1 +
					y_2 \bar{y}_2 \right) \left(
					I_{12} + \bar{I}_{12} + h +
					\bar{h}\right)
					\right. \right. \nonumber \\ &
					& \qquad\qquad \left.\left.-
					y_1 \bar{y}_2 \left( I_{22} +
					\kappa \bar{I}_{11} \right) -
					\bar{y}_1 y_2 \left(
					\bar{I}_{22} + \kappa I_{11}
					\right) \right] +
					O(\varepsilon^2)\right. \! 
					\bigg\}\nonumber
					\equiv  e^{\varphi_c^0} (1 +
					\varepsilon \, \chi +O(\varepsilon^2)).
\end{eqnarray}
Eq.(\ref{beta}) gives the value of $\beta_4$ to first order
$\beta_4=\varepsilon \beta$. Recalling the expression of the
unperturbed conformal factor $e^{\varphi_c^0}$ with only three sources
we have
\begin{equation}\label{betaderivvarphi}
\beta_4 = -4\varepsilon\, {\left. e^{\varphi_c^0/2}\,\partial_z\,
	e^{-\varphi_c^0/2} \right|}_{z=t} = 2\varepsilon \, {\left. \partial_z
	\varphi_c^0(z) \right|}_{z=t}~.
\end{equation}
The above obtained relation can be understood by expanding Liouville
equation around the unperturbed solution.

We can exploit such a result and Polyakov relation to compute to order
$\varepsilon$ the classical action for the new solution
\begin{equation}
\frac{\partial S_{cl}[\eta_1,\eta_2,\eta_3,\varepsilon]}{\partial t} =
-\frac{\beta_4}{2} = -\varepsilon \frac{\partial \varphi_c^0}{\partial t}
\end{equation}
and using again eq.(\ref{Xrelation}) 
we reach for the semiclassical four point function with small $\alpha_4$
\begin{equation}\label{4pointfunctionspecial}
\left< V_{\alpha_1}(0)  V_{\alpha_2}(1)
	V_{\alpha_3}(\infty) V_{\alpha_4}(t)\right>_{sc} =
	\left<V_{\alpha_1}(0) V_{\alpha_2}(1)
	V_{\alpha_3}(\infty)\right>_{sc} ~e^{2\alpha_4 \phi_c^0(t)}.
\end{equation}
It is easily checked that the four point function
(\ref{4pointfunctionspecial}) has the correct transformation properties with
dimensions 
$\alpha_4/b$ for the vertex field  $V_{\alpha_4}(z_4)$ in agreement
with the semiclassical dimensions $\alpha_4(1/b-\alpha_4)$ keeping in
mind that we have been working to first order in $\alpha_4$, and thus
we can write to first order in $\alpha_4$
\begin{equation}\label{4pointfunction}
\left< V_{\alpha_1}(z_1)  V_{\alpha_2}(z_2)
	V_{\alpha_3}(z_3) V_{\alpha_4}(z_4)\right>_{sc} =
	\left<V_{\alpha_1}(z_1) V_{\alpha_2}(z_2)
	V_{\alpha_3}(z_3)\right>_{sc} ~e^{2\alpha_4 \phi_c^0(z_4)}~.
\end{equation}

\section{Generalization to $n$-point functions}

We can generalize some of the results obtained above to $n$ arbitrary
sources and $m$ infinitesimal sources. 
The discussion we have been 
performing in the case of three sources which leads to the
inhomogeneous equation (\ref{inhomogeneous}) remains valid also in
this case; the 
only difference is that now we do not know the explicit form of the
unperturbed solutions $y_1,y_2$. 
The accessory parameter $\beta_t =
\varepsilon\beta$ is again given by eq.(\ref{betaderivvarphi})
\begin{equation}\label{eq:beta-generale}
\beta = 2 {\left. \partial_z \, \varphi_c^0(z) \right|}_{z=t}
\end{equation}
where now $\varphi_c^0$ is the conformal field which solves the
problem in presence of the $n$ finite sources. Thus we have a general
relation between the value of the accessory parameter relative to the
infinitesimal source in $t$ and the conformal factor for the
unperturbed background and thus we can extend the result
(\ref{4pointfunction}) to 
$n$ finite sources plus an infinitesimal one. Finally due to the
additive nature of the perturbation with $m$ infinitesimal sources 
we have for the $n+m$ semiclassical correlation function
\begin{equation}
\left< V_{\alpha_1}(z_1)\dots V_{\alpha_n}(z_n)\, V_{\gamma_1}(t_1)
\dots V_{\gamma_m}(t_m) \right>_{sc} = \left< V_{\alpha_1}(z_1)\dots
V_{\alpha_n}(z_n)\right>_{sc} \prod_{j=0}^{m} e^{2\gamma_j
\phi_c^0(t_j)}.
\end{equation}
Such relations were already argued in \cite{ZZsphere}.

\section{The Green function on the sphere with three singularities}
\label{green}\label{greensection}

From the above derived results we can extract the exact Green function
on the sphere in presence of three finite singularities. The equation
for the Green function is
\begin{equation}\label{greenequation}
-\Delta \, g(z,t) + 8\pi\mu b^2 e^{\varphi_B(z)} \, g(z,t) = 2\pi \,
 \delta^2(z-t)
\end{equation}
where $\varphi_B$ is the classical solution in presence of three
finite singularities.
Such a Green function can be computed from the result obtained in
Sect.3. In fact we have found a solution to
\begin{equation}\label{eq:sorgente-infinitesima-2}
- \Delta \varphi + 8\pi\mu b^2 \, e^{\varphi} = 8\pi \sum_{i=1}^3
  \eta_i \, \delta^2(z-z_i) +8\pi\varepsilon \delta^2(z-t)
\end{equation}
for infinitesimal $\varepsilon$ i.e. $\varphi=\varphi_B+
\varepsilon \chi$. Substituting we obtain
\begin{equation}
-\Delta \chi + 8\pi\mu b^2 e^{\varphi_B}\, \chi = 8\pi\delta^2(z-t)
\end{equation}
i.e. we have $\displaystyle{g(z,t) = \frac{\chi}{4}}$. From
eq.(\ref{expphi}) we have 
\begin{eqnarray}\label{greenfunction}
g(z,t) & = & - \frac{1}{2 w_{12} \left[\kappa y_1(z) \bar{y}_1(\bar{z}) -
	y_2(z) \bar{y}_2(\bar{z})\right]} \Big\{ \left[\kappa y_1(z)
	\bar{y}_1(\bar{z}) + y_2(z) \bar{y}_2(\bar{z}) \right] \,
	\cdot \nonumber \\ & & \qquad \qquad \cdot \left[ I_{12}(z,t)
	+ \bar{I}_{12}(\bar{z},\bar{t}) + h(t) +
	\bar{h}(\bar{t})\right] \nonumber \\ & & \qquad\qquad - y_1(z)
	\bar{y}_2(\bar{z}) \left[ I_{22}(z,t) + \kappa
	\bar{I}_{11}(\bar{z},\bar{t}) \right] \nonumber \\ & & \qquad
	\qquad - \bar{y}_1(\bar{z}) y_2(z) \left[
	\bar{I}_{22}(\bar{z},\bar{t}) + \kappa I_{11}(z,t) \right]
	\Big\}.
\end{eqnarray}
It is possible to verify directly that (\ref{greenfunction}) satisfies
eq.(\ref{greenequation}) and is regular on the three finite sources. 
Actually expression (\ref{greenfunction}) is completely general, i.e. it
applies also for the case of a background given by $n$ finite sources
with $y_i$ solutions of the related fuchsian equation. In the case of
$n=3$ we know the explicit form of $y_i$.
One would expect the Green function $g(z,t)$ to be symmetric in the
arguments. This is far from evident from the expression
(\ref{greenfunction}). The 
differential operator $D= -\Delta_{LB}+1$ is hermitean in the
background metric $e^{\varphi_B}d^2z$. As a result also its inverse $G
= D^{-1}$ is hermitean $G=G^+$. $G$ is represented by $g(x,t)$ which
is also real and thus we have $g(z,t) = g(t,z)$.

\section{The quantum determinant}

The complete action is given by eqs.(\ref{classicalaction}) and
(\ref{quantumaction}) and the quantum $n$-point function by
\begin{equation} 
\left< V_{\alpha_1}(z_1)  V_{\alpha_2}(z_2) \dots
V_{\alpha_n}(z_n)\right>= e^{-S_{cl}[\phi_B]} \int D[\chi]~ e^{-S_q}. 
\end{equation}
We recall that $S_{cl}$ is $O(1/b^2)$ 
while the first integral appearing in the quantum action
(\ref{quantumaction}) can be expanded as 
\begin{equation}
\frac{1}{4\pi}\int_\Gamma\left((\partial_a
\chi)^2 + 8\pi \mu b^2 e^{\varphi_B}\chi^2 + 8\pi \mu b^2
e^{\varphi_B}(\frac{4b \chi^3}{3!}+\frac{8 b^2\chi^4}{4!}+\dots\right) d^2z . 
\end{equation}
From now on we shall denote by $\varphi_B$ the classical solution with
three singularities at $z_1,z_2,z_3$ and with charges
$\eta_1,\eta_2,\eta_3$. 
In performing the perturbative expansion in $b$ we have to keep the
$\eta_1,\eta_2,\eta_3$ constant \cite{ZZsphere}. 
The $O(b^0)$ contribution to the three point function is given by
\begin{equation}\label{determinant}
({\rm Det}D)^{-\frac{1}{2}}= \int D[\chi] e^{-\int \chi(z) D \chi(z)f(z)d^2z}
\end{equation}
where $f(z) = 8\pi\mu b^2 e^{\varphi_B(z)}$ and 
$
D=(-\Delta_{LB}+1)/4 \pi
$
being $\Delta_{LB}=f^{-1} \Delta$ the Laplace-Beltrami operator on the
background $f(z)$ of the three charges. It provides the one loop
quantum correction to the semiclassical results we have been
discussing above.

When confronted to the computation of a functional
determinant the first idea is to use the $Z$-function regularization;
but that does not work because such a regularization is invariant
under conformal transformations and as such leaves the dimensions of
the vertex functions at their semiclassical values; in particular the
weights of the cosmological term remain $(1-b^2,1-b^2)$ instead of
becoming $(1,1)$
and further quantum corrections which are higher order in $b^2$ cannot
mend such a 
discrepancy. The situation is similar to the one discussed by D'Hoker,
Freedman and 
Jackiw \cite{DJ,DFJ} and the one considered by Takhtajan \cite{takhtajan}
with the use of an  
invariant regularization of the Green function (Hadamard
regularization) and $Q=1/b$. 
Thus a different way to define the regularized determinant has to be
devised. One can take the derivative of the logarithm of the
determinant thus 
exposing the role of the regularized Green function at coincident
points. 
We have  
\begin{equation}\label{etaderivative}
\frac{\partial}{\partial \eta_j}\left(\log ({\rm Det}D)^{-\frac{1}{2}}\right)=
-2\mu b^2 \int\frac{\partial\varphi_B}{\partial\eta_j}(z) g(z,z)
  e^{\varphi_B(z)} d^2z.
\end{equation}
In the above equation the Green function at coincident points appears
and such a quantity has to be regularized. We have already seen that
the invariant regularization gives rise to a theory in which
the cosmological term $e^{2b\phi(z)}$ does not have weights $(1,1)$ and as
such does not give rise to a theory invariant under the whole
(infinite dimensional) conformal group.

\noindent
We shall adopt here the regularization proposed by Zamolodchikov and
Zamolodchikov \cite{ZZpseudosphere} (ZZ regulator) for  
perturbative calculations on the pseudosphere i.e.
\begin{equation}
g(z,z) = \lim_{z'\rightarrow z}\left(g(z,z') + \log|z-z'|\right).
\end{equation}  
As under an $SL(2,C)$ transformation
$
w=(az+b)/(cz+d)
$
the Green function is invariant in value $ g^w(w,w') = g(z,z')$
and we have
\begin{equation}
g^w(w,w) = g(z,z) +\log\left|\frac{\partial w}{\partial z}\right| = g(z,z)
+\log\left|\frac{1}{(cz+d)^2}\right|.  
\end{equation}
Using the above relation one can compute the change of
(\ref{etaderivative}) under $SL(2,C)$ transformations.
Taking into account the contribution of the boundary term we have for
$j=1,2,3$, $k=2,3,1$, $l=3,1,2$ 
\begin{equation}
\frac{\partial}{\partial \eta_j}\log(({\rm Det}D)^{-\frac{1}{2}}
-\frac{\partial 
X_\infty}{\partial \eta_j}=
f_j(\eta_1,\eta_2\eta_3)
-2\log\left|\frac{(z_j-z_k)(z_j-z_l)}{(z_k-z_l)}\right|. 
\end{equation}
Integration of the above equation gives
\begin{equation}
c(\eta_1,\eta_2\eta_3) - 2(\eta_1+\eta_2-\eta_3)\log|z_1-z_2|
- 2(\eta_2+\eta_3-\eta_1)\log|z_2-z_3|
- 2(\eta_3+\eta_1-\eta_2)\log|z_3-z_1|
\end{equation}
as $O(b^0)$ correction i.e. the one loop correction,
and we have obtained  the three-point function with the correct
quantum dimensions 
$\Delta_j = \eta_j(1-\eta_j)/b^2 + \eta_j$. Thus the
situation is very similar to what happens on the pseudosphere, where
the one loop corrections with the ZZ regulator provide the exact
quantum dimensions \cite{ZZpseudosphere,MT2}.

\section{Two loop contributions}

The graphs contributing to two loop are shown in Figure.1 . Of them
graph (c) is convergent and invariant under under $SL(2,C)$; (a) and
(b) are not due to the appearance of the regularized Green function
$g(z,z)$, while (d) and (e) arise from the boundary term. Integrating by
parts a number of times and using the equation for the Green function
one can prove that the sum is invariant under translations,
dilatations and inversion and thus under the whole global conformal
group \cite{MV}. This shows that the dimension are not altered at two
loop. One also envisages a general procedure to higher loop but that
has not yet been done explicitly; on the pseudosphere one can give an
explicit procedure to all orders \cite{MTx}.

\begin{figure}[h]
\includegraphics[width=18pc]{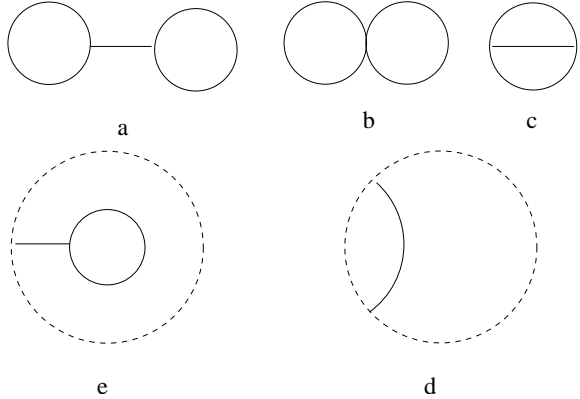}\hspace{2pc}%
\begin{minipage}[b]{18pc}\caption{\label{label}Two loop contributions}
\end{minipage}
\end{figure}

\section{The pseudosphere}
In this section we shall apply the developed technique to the
pseudosphere \cite{MT3}. 
For the one point function, this analysis goes well beyond the
previous perturbative expansions performed in \cite{ZZpseudosphere,
MT1, MT2} where $\alpha$ has been taken small;
in fact our result corresponds to the summation of an infinite
class of perturbative graphs. Thus, we obtain a strong check of
the ZZ bootstrap formula for the one point function
\cite{ZZpseudosphere}, which includes all the previous perturbative
checks.
\noindent We start from the Liouville action on the pseudosphere
in presence of $N$ sources characterized by heavy charges
$\eta_1,\dots,\eta_N$, given in \cite{MT2}. The standard
representations of the pseudosphere are the unit disk $\Delta$ and the
upper half plane $H$. Here we shall mostly use the $\Delta$
representation.
Decomposing the Liouville field as before 
the Liouville action separates into a classical part, depending
only on the background field
$\varphi_{\scriptscriptstyle\hspace{-.05cm}B}$, and a quantum
action for the quantum field $\chi$,
$
S_{\Delta,\,N}[ \,\phi\,]  \,=\,S_{cl}[
\,\varphi_{\scriptscriptstyle\hspace{-.05cm}B}\,] +
S_{q}[\,\varphi_{\scriptscriptstyle\hspace{-.05cm}B},\,\chi\,]
$
which have expressions similar to the ones appearing in the case of
the sphere \cite{MT2,MT3}.
The coupling constant $b$ is still related
to the parameter $Q$ occurring in the central charge $c=1+6Q^2$ by
$Q=1/b+b$ \cite{CT}. 
Again at semiclassical level, we have
\begin{equation}\label{semiclassic N point}
\left\langle \, V_{\alpha_1}(z_1)\dots V_{\alpha_N}(z_N)
\,\right\rangle_{sc}\,=\,
\frac{e^{-S_{cl}(\eta_1,\,z_1;\dots;\,\eta_N,\,z_N)}}{e^{-S_{cl}(0)}}
\end{equation}
where $S_{cl}(\eta_1,z_1;\dots;\eta_N,z_N)$ is the classical
action $S_{cl}[ \,\varphi_{\scriptscriptstyle\hspace{-.05cm}B}\,]$
computed on the solution
$\varphi_{\scriptscriptstyle\hspace{-.05cm}B}$ of the Liouville
equation with sources. 
One can see that the transformation law of $S_{cl}[
\,\varphi_{\scriptscriptstyle\hspace{-.05cm}B}\,]$ assigns to the
vertex operator $V_\alpha(z)$ the semiclassical dimensions
$\alpha\,(1/b-\alpha)=\eta\,(1-\eta)/b^2$  \cite{MT2} as already found
in Sect.(\ref{clt}) on the sphere.

\noindent For the one point function, we have a single heavy
charge $\eta_1=\eta$, which can be placed in $z_1=0$, and the
explicit solution of the Liouville equation is
$\varphi_{\scriptscriptstyle\hspace{-.05cm}B}=\varphi_{cl}$, given
by \cite{seiberg}
\begin{equation}\label{phiclassic}
    e^{\varphi_{cl}}\,=\,\frac{1}{\rule{0pt}{.4cm}\pi\mu b^2}\;
    \frac{(1-2\eta)^2}{\big[\,\rule{0pt}{.4cm}(z\bar{z}
\hspace{.03cm})^{\eta}-(z\bar{z}\hspace{.03cm})^{1-\eta}\,\big]^2}\;.
\end{equation}
The classical action computed on
this background gives the semiclassical one point function
\begin{equation}\label{one point classical term}
\left\langle \, V_{\eta/b}(0) \,\right\rangle_{sc} \,=\,
\exp\left\{-\,\frac{1}{b^2}\;\Big(\,\eta\,\log\left[\,\pi
        b^2\mu\,\right]+2\eta+(1-2\eta)\,\log(1-2\eta)\,\Big)\,\right\}\;.
\end{equation}
Again to go beyond this approximation we need the Green
function on the background field given by (\ref{phiclassic}) and,
to do this, we employ the method explained in
Sect.(\ref{greensection}). 
The only distinguishing feature is that as we work on the disk, we
have to impose the Cardy condition \cite{Cardy} and the regularity
condition at infinity on the classical energy momentum tensor. One
can express them more easily in the upper half plane
${H}=\{\,\xi\in {C}\,;\, \textrm{Im}(\xi)>0\,\}$
representation 
where they read
$\widetilde{Q}(\xi)\,=\,\overline{\rule{0pt}{.45cm}\widetilde{Q}}(\xi)$
and $\xi^4\,\widetilde{Q}(\xi) \,\sim\,
O(1)$ when $\xi\rightarrow\infty$, respectively.\\
Writing the classical field in presence of two sources of
charges $\eta$ in $0$ and $\varepsilon$ in $t$ taken real, as an
expansion up to the first 
order in $\varepsilon$, i.e.
\begin{equation}\label{varphi2 eps expansion}
    \varphi_{2}(z)\,=\,\varphi_{cl}(z)+ \epsilon\,
    \chi(z,t\hspace{.04cm})+ O(\epsilon^2\hspace{.03cm})
\end{equation}
one finds that this analysis leads to the following expression for
$\chi(z,t\hspace{.04cm})$
\begin{eqnarray}\label{psi(z,t)}
 \chi(z,t\hspace{.04cm}) \hspace{-.1cm}& = &\hspace{-.1cm} -\,
\frac{2}{w_{12}\,(y_1 \bar{y}_1-y_2\bar{y}_2)}\,
 \left\{\,
\big(\,y_1\bar{y}_1+y_2\bar{y}_2\,\big)\,
\big(\,I_{12}+\bar{I}_{12}+2\,h_0\,\big)\rule{0pt}{.4cm}\right.
\\
\rule{0pt}{.5cm} & & \hspace{5.3cm}\left.
\rule{0pt}{.4cm}-\;\bar{y}_1 y_2\,I_{11}-\,y_1\bar{y}_2\,I_{22}
 -\, y_1\bar{y}_2
\,\bar{I}_{11}-\,\bar{y}_1 y_2\,\bar{I}_{22} \,\right\}\nonumber
\end{eqnarray}
where $w_{12}=y_1y_2'-y_1'y_2=1-2\eta$ is the constant wronskian,
the $I_{ij}$ are defined similarly to what done for the sphere
and $h_0$ is a free real parameter which cannot be determined
through monodromy arguments because it is the coefficient of a
solution of the homogeneous equation. It is fixed by requiring the
vanishing of $\chi(z,t\hspace{.04cm})$ at infinity, i.e. when
$|z|\rightarrow 1$, in order to respect the boundary condition 
on the pseudosphere.
The Green function on the background $\varphi_{cl}(z)$ is given by
$g(z,t\hspace{.04cm})=\chi(z,t)/4$. By exploiting
the invariance under rotation, we can write our result for a
generic complex $t\in \Delta$. The final expression of the exact Green
function in the explicit symmetric form is
\begin{eqnarray}\label{g(z,t) symmetric}
 \rule{0pt}{1cm}g(\hspace{.02cm}z,t\hspace{.04cm}) & = &
 -\,\frac{1}{\rule{0pt}{.4cm}2}\;
\frac{1+(z\bar{z}\hspace{.04cm})^{1-2\eta}}
{\rule{0pt}{.4cm}1-(z\bar{z}\hspace{.04cm})^{1-2\eta}}\;
\frac{1+(\hspace{.02cm}t\bar{t}\hspace{.04cm})^{1-2\eta}}
{\rule{0pt}{.4cm}1-(\hspace{.02cm}t\bar{t}\hspace{.04cm})^{1-2\eta}}\;
\log\omega(\hspace{.02cm}z,t\hspace{.04cm})\,
-\,\frac{1}{\rule{0pt}{.4cm}1-2\eta}\,
\\
\rule{0pt}{1cm} & & \hspace{-2cm}-\,
\frac{1}{\rule{0pt}{.4cm}\,1-(z\bar{z}\hspace{.04cm})^{1-2\eta}}\;
\frac{1}{\rule{0pt}{.4cm}\,1-(\hspace{.02cm}t\bar{t}
\hspace{.04cm})^{1-2\eta}}\;\left\{\,
(z\bar{t}\hspace{.05cm})^{1-2\eta}\,\Big(\,B_{\,z/\,t}
\big(\,2\eta,\,0\,\big)-B_{\,z\bar{t}\,}\big(\,2\eta,\,0\,\big)\,\Big)
\rule{0pt}{.6cm}\right.
\nonumber \\
& &\rule{0pt}{.6cm} \hspace{4cm}\left.
+\,(\bar{z}t\hspace{.02cm})^{1-2\eta}\,\Big(\,B_{\,t/z}
\big(\,2\eta,\,0\,\big)-B_{\,1/(z\bar{t}\hspace{.02cm})\,}
\big(\,2\eta,\,0\,\big)\,\Big)
+\textrm{c.c.}\rule{0pt}{.6cm}\;\right\} \nonumber
\end{eqnarray}
where $\omega(\hspace{.02cm}z,t\hspace{.04cm})$ is the $SU(1,1)$
invariant ratio
\begin{equation}\label{SU(1,1) invariant ratio}
\omega(\hspace{.02cm}z,t\hspace{.04cm})\,=\,\frac{(z-t)\,
(\bar{z}-\bar{t}\hspace{.04cm})}
{\rule{0pt}{.4cm}\,(\hspace{.02cm}1-z\bar{t}\hspace{.04cm})\,
(\hspace{.02cm}1-\bar{z}t\hspace{.03cm})}
\end{equation}
and $B_x(a,0)$ is a particular case of the incomplete Beta
function $B_x(a,b)$
\begin{equation}
    B_{x}(\hspace{.02cm}a,0\hspace{.03cm})\,=\,
\frac{x^a}{a}\,F(\hspace{.02cm}a,1;a+1;\,x)\,=\,\int_0^{\,x}
    \frac{y^{a-1}}{1-y}\;dy\,=\,\sum_{n\,=\,0}^{+\infty}
\frac{x^{a+\,n}}{a+n}\;.
\end{equation}
Moreover, in the limit $\eta\rightarrow 0$, we recover the
propagator on the pseudosphere
without sources given in \cite{DFJ, ZZpseudosphere}.
\noindent From (\ref{g(z,t) symmetric}), we find
\begin{eqnarray}\label{g(z,z)}
    g(z,z)\,&\equiv &\lim_{t\,\rightarrow\,
    z}\,\left\{\,g(z,t\hspace{.06cm})+\frac{1}{2}\,\log\left|\,z-t\,
\right|^2\,\right\}\;\\
& = &
\Bigg(\,\frac{1+(z\bar{z}\hspace{.04cm})^{1-2\eta}}
{\rule{0pt}{.4cm}1-(z\bar{z}\hspace{.04cm})^{1-2\eta}}\,\Bigg)^2
\log\big(\,1-z\bar{z}\,\big)\,
-\,\frac{1}{\rule{0pt}{.4cm}1-2\eta}\;\frac{1+(z\bar{z}
\hspace{.04cm})^{1-2\eta}}{\rule{0pt}{.4cm}1-(z\bar{z}
\hspace{.04cm})^{1-2\eta}}
\\
\rule{0pt}{1cm}
 & &   +\,\frac{2\,(z\bar{z}\hspace{.04cm})^{1-2\eta}}{
\big(\rule{0pt}{.45cm}\,1-(z\bar{z}\hspace{.04cm})^{1-2\eta}\,\big)^2}\,
\left(
B_{z\bar{z}}\big(\,2\eta\,,0\,\big)+B_{z\bar{z}}\big(\,2-2\eta\,,0\,\big)
\rule{0pt}{.5cm}\right.\nonumber\\
& &\hspace{6.2cm} \left.\rule{0pt}{.5cm}+
2\gamma_E+\psi(2\eta)+\psi(2-2\eta)-\log z\bar{z}
\,\right)\nonumber
\end{eqnarray}
where $\gamma_{E}$ is the Euler constant and
$\psi(x)=\Gamma'(x)/\Gamma(x)$.

\noindent In the case of the one point function, expression
(\ref{etaderivative}) can be explicitly computed and the result is
\begin{equation}\label{log det integral}
\frac{\partial}{\partial \eta}\,\log
\big(\,\textrm{Det}\,D(\eta,0)\,\big)^{-1/2}
=\;2\,\gamma_{\,\scriptscriptstyle\hspace{-.05cm}E}\, +\,2\,
\psi(1-2\eta)\,+\,\frac{3}{1-2\eta}\;.
\end{equation}
Integrating back in $\eta$ with the initial condition given in
\cite{ZZpseudosphere}, i.e. $
\left.\big(\,\textrm{Det}\,D(\eta,0)\,\big)^{-1/2}\,\right|_{\,\eta\,=\,0}
\hspace{-.2cm}=\,1$, we find
\begin{equation}\label{quantum det eta}
\log \big(\,\textrm{Det}\,D(\eta,0)\,\big)^{-1/2}\,=\,
2\,\gamma_{\,\scriptscriptstyle\hspace{-.05cm}E} \,\eta-\log
\Gamma(1-2\eta) -\frac{3}{2}\,\log(1-2\eta)\;.
\end{equation}
Putting this result together with the classical contribution
(\ref{one point classical term}), we have
\begin{eqnarray}\label{classandoneloop}
\rule{0pt}{.6cm} \log \left\langle \, V_{\eta/b}(0)
\,\right\rangle  \hspace{-.1cm} & = & \hspace{-.2cm}
-\,\frac{1}{b^2}\;\Big(\,\eta\,\log\left[\,\pi
        b^2\mu\,\right]+2\eta+(1-2\eta)\,\log(1-2\eta)\,\Big)
        \nonumber\\
 \rule{0pt}{.8cm} & & +\,\Big(\,
2\,\gamma_{\,\scriptscriptstyle\hspace{-.05cm}E} \,\eta-\log
\Gamma(1-2\eta) -\frac{3}{2}\,\log(1-2\eta)\,\Big) + O(b^2)
\end{eqnarray}
to all orders in $\eta$.
We can compare (\ref{classandoneloop}) with the result obtained
by ZZ within the bootstrap approach \cite{ZZpseudosphere}
\begin{equation}
 \label{one-point function}
\left\langle \,V_{\alpha}(z_1) \,\right\rangle =
\frac{U(\alpha)}{(\,1-z_1\bar{z}_1\,)^{\,2\alpha(Q-\alpha)}}
\end{equation}
where $U(\alpha)$ has been determined through the bootstrap method
\cite{ZZpseudosphere} with the result for the basic vacuum
\begin{equation}
U(\alpha)\,=\,U_{1,1}(\alpha)\,=\,\big(\,\pi\mu\gamma(b^2)\,\big)^{-\alpha/b}
\frac{\Gamma(Q\,b)\,\Gamma(Q/b)\,Q}{\rule{0pt}{.36cm}
\Gamma\big((Q-2\alpha)\,b\big)\,\Gamma\big((Q-2\alpha)/b\big)\,(Q-2\alpha)}
\end{equation}
where $\gamma(x)=\Gamma(x)/\,\Gamma(1-x)$.
Our result (\ref{classandoneloop}) agrees with the expansion in
$b^2$ of $U(\eta/b)$ and it
corresponds to the summation of an infinite class of graphs of the
usual perturbative expansion \cite{ZZpseudosphere, MT1,
MT2}. With some more work one could compute the two loop
correction to eq.(\ref{classandoneloop}).
The technique developed above can be applied to compute
the two point function with one arbitrary charge $\eta$ and
another charge $\varepsilon$ to first order in $\varepsilon$ \cite{MT3}
and compared successfully with the results of the conformal bootstrap
approach.

\section{Conclusions}
We have pursued the functional approach to Liouville quantum field
theory, in the usual meaning of computing a classical stable
background and integrating over the quantum fluctuations around it.
We found that the standard technique of regulating the quantum
determinants ($Z$-function) violates conformal invariance. A non
conventional technique i.e. the ZZ regularization is necessary both in
computing the functional determinant and higher order graphs.  The
boundary terms term play an essential role in the proof of the
invariance at one and two loop order.  The exact Green function on the
background of the the sphere with three finite singularities has been
obtained in terms of quadratures. The explicit form of the Green
function on the background of the pseudosphere in presence of a
singularity has been given and the exact quantum
determinant computed. In corresponds to the resummation of an infinite
family of graphs and the results agree with the conformal bootstrap results of
A.B. Zamolodchikov and Al.B. Zamolodchikov. Applications to boundary
Liouville theory on the disk and to higher loop are underway \cite{MTx}.

\ack

I am grateful to Erik Tonni for collaboration on part of the work
described here and for useful discussions.


\eject

\end{document}